# An Investigation of the Relationship Between Crime Rate and Police Compensation

Jhancy Amarsingh[1], Likhith Kumar Reddy Appakondreddigari[1], Ashish Nunna[1], Charishma Choudary Tummala[1], John Winship[1], Alex Zhou, and Huthaifa I. Ashqar[1,2]

[1] University of Maryland, Baltimore County
[2] Arab American University

**Abstract**

The goal of this paper is to assess whether there is any correlation between police salaries and crime rates. Using public data sources that contain Baltimore Crime Rates and Baltimore Police Department (BPD) salary information from 2011 to 2021, our research uses a variety of techniques to capture and measure any correlation between the two. Based on that correlation, the paper then uses established social theories to make recommendations on how this data can potentially be used by State Leadership. Our initial results show a negative correlation between salary/compensation levels and crime rates.

**Introduction**

Baltimore is experiencing an increase in crime rates, to the point where it is becoming a major social issue. To combat this trend, State Leadership wants to increase Police Department budgets, part of which will be an increase in police officer salaries. As a means to help inform this process, and create a viable police salary model that balances public safety with public expense, our research focuses on whether there is any correlation between police salaries and crime rates over time. And, if there is a correlation, how can that correlation be used to affect public policy?

Our team leveraged the datasets made available via the Open Baltimore initiative, to include crime rate and police officer salary data from 2011 to 2021. Using a variety of modeling techniques, we were able to assess and measure a negative correlation between the crime rates and police salary.

Using this information, the paper provides some further conclusions on how to potentially apply these findings to drive public policy. We use social theories established by Becker[1], Sherman[2] and others to make recommendations for further analysis, and to provide an initial concept for how to translate the correlation between police salaries and crime rates into actionable public policy.

**Research Question:**
The fundamental question we are trying to answer is whether or not there is any correlation between police salaries and the level of crime over time. To do that, we break the problem down into three discrete steps.

### *Step 1: Baseline the level of violent crime in Baltimore from 2011 to 2021:*
The first thing to quantify is the crime rates in Baltimore over time. Using detailed crime data from the BPD (see next section for a more thorough description), our team summarized the 250,000 plus crime entries into a month-to-month crime rate time series from 2011 to 2021.

Important within this analysis is the ability to account for fluctuations in crime rates month-to-month. For example, the data showed that there are more crimes committed in the summer than winter. Our analysis took this into account, and attempted to smooth the crime rate over a given 12 month period. Further fluctuations occur based on 'macro' events, the most notable within this timeframe being the Freddy Grey case in 2015. Here we leveraged work done by Kolodrubetz[3] to account for the 'Freddy Grey effect.'

In parallel, we performed a spot comparison to another city (in this case San Francisco) to ensure that the Baltimore Crime rates are 'reasonable.' In this context, reasonable equates to roughly the same levels of crime per capita, in particular violent crime. This comparison is to ensure Baltimore is not a truly unique case (i.e. a modern 'wild west') and that broad social theories can be legitimately applied. This is more important for the later analysis regarding public policy.

### *Step 2: Create a baseline of police salaries over the same time frame:*
This next step is to establish the salary/compensation side of the equation, and was a relatively straightforward process of using publicly available police officer salary and gross pay data from 2011 to 2021. That said, there is some additional salary / compensation analysis required to adjust for things such as the mix of seniority levels techniques to adjust for inflation.

### *Step 3: Compare the two using several different methodologies, and quantify any correlation*
This process is described later in the Methods, Analysis and Results Sections

### *Step 4: If there is a correlation, how can that correlation be used to affect public*

*policy?*

Based on the results, our analysis provides recommendations on whether or not it's worth investing in additional analysis to (a) refine the results, and (b) translate them into potential public policy. If so, how could the results be combined with existing theoretical social theories to help decision makers to maximize the public good with limited resources.

## Literature Review

Our literature review covered two main areas: crime rate vs public expense, and the application of modeling techniques to inform police procedures and policy questions.

For the first area, beyond understanding what work has been done to relate crime rates to salaries or other 'cost-based' metrics, an additional goal was to establish a means to translate our findings into actual public policy. In other words, ensure that any connection we might make between our two main variables (salaries and crime rates) could actually be useful, according to established research.

To that end, and on the recommendation of a classmate, we started with Gary Becker[1] whose work applying economic theory to the subject of crime is well established. We will revisit some of the specifics in the Conclusion Section, but generally speaking Becker's work does establish a connection between police compensation, crime rates, and the idea of social loss.

Becker's work is very high-level and 'macro,' and so we also reviewed several more low-level, recent studies on crime-policy. A major resource in this effort was Sherman et. al.[2] which captures "the major conclusions of a 1997 report to Congress, which was based on a systematic review of more than 500 scientific evaluations of crime prevention practices. This Research in Brief summarizes the research methods and conclusions found in that report."

Specific 'manpower' related findings are from Press[4] and Chaiken[5] which found that extra police patrols in high crime hot spots reduce crime in those places. Another study (Abrahamse et. al. [6]) confirmed that specialized 'repeat offender units' were effective at reducing the impact of "high risk repeat offenders by monitoring them and returning them to prison more quickly than when they are not monitored to reduce their crimes." All of these studies point to the idea that a larger police presence, to include highly trained and experienced officers, increases public safety

Additional inputs came from McCollister et. al.[7], and their work estimating the cost to society of individual crimes is essential to the economic evaluation of many social programs, such as substance abuse treatment and community policing. A review of the crime-costing literature reveals multiple sources, including published articles and government reports, which collectively represent the alternative approaches for estimating the economic losses associated with criminal activity.

> *"We cannot defund the police. We need to re-fund the police. Instead of less funding, we need more investment in public safety..." Governor Larry Hogan[8]*

The quote above from 15 Oct 21 brings these broad issues home to Baltimore, but the question is whether or not the current stand on police funding is the correct approach? It is this question we are trying to answer. To help with that, in May, 2020 the Baltimore Police Department began a significant upgrade to its new Records Management Systems to allow the department to transition from a paper-based system into a fully digital reporting environment.[9] Violent crime is a well known social problem affecting both the quality of life and the economical development of a society. Its prediction and prevention is, therefore, an important asset for law enforcement agencies. We can say that this indeed needs more manpower and economic might.

A related and exciting development within this domain is 'Predictive Policing,' which refers to gathering loads of data and applying algorithms to deduce where and when crimes are most likely to occur. There are a few such applications being used by police departments across the United States. In particular, "PredPol[10]" used by the Los Angeles and Atlanta Police Departments uses place, time and type of crime to create hotspot maps that help police to decide nightly patrol routes. Another Machine Learning tool, HunchLab, used by the New York & Miami Police Departments uses social and behavioral analysis to generate predictions. According to LAPD, a significant percentage drop in burglaries is reported after deployment of machine learning tools.

In this work, we tackle both aspects: correlation between funds to law enforcement and crime rate and how that can be inferred. We propose to predict the impact of having higher funds to law enforcement agencies on crime rate with regards to historic data on both crime rates and salaries to police personnel using regression and optimize the distribution of funds to police officers, taking into account the previous predictions.

## Datasets

A dataset is a collection of related sets of information that is composed of separate elements but can be manipulated as a unit by a computer. In the case of tabular data, a data set corresponds to one or more database tables, where every column of a table represents a particular variable, and each row corresponds to a given record of the data set in question. It not only serves as a major data storage format, but also as a standard container for the results of most algorithms. Data sets can hold information such as medical records or insurance records, to be used by a program running on the system. Data sets are also used to store information needed by applications or the operating system itself, such as source programs, macro libraries, or system variables or parameters. A dataset, in its most basic form, only contains data in the form of a matrix of numerical values. A Dataset has many advantages

- High-level abstraction and custom view into structured data.
- Ease-of-use of APIs with structure.
- Performance and Optimization.

As mentioned above, our research leverages two main datasets: Crime Rate data and Salary data. These are shown in Figures 1 and 2 below.

Figure 1: Crime Rate - City of Baltimore Crime Data from 2011-2021

Figure 2: Salary/Gross Pay Data for Baltimore Employees, specific to the Police Department

## Methods

Pandas, Python's core data analysis support library, provides fast, flexible, and explicit data structures designed to handle relational, tokenized data in a simple and intuitive way.

Pandas is suitable for working with the following types of data: tabular data with heterogeneous columns, similar to SQL or Excel tables; ordered and unordered (non-fixed frequency) time series data; Matrix data with row labels, including isomorphic or heterogeneous data; and any other form of observations, statistical data sets, which do not need to be pre-tagged when transferred to the Pandas data structure.

Pandas main data structures are Series (one-dimensional data) and DataFrame (two-dimensional data), which are sufficient to handle most typical use cases in finance, statistics, social sciences, engineering, and other fields. For R users, DataFrame provides richer functionality than the R language data.frame. Pandas is based on NumPy and can be easily integrated with other third-party scientific computing support libraries.

Pandas has several key advantages:

- Handling missing data in floating-point and non-floating-point data, represented as NaN.
- Variable size: insert or delete columns of multidimensional objects such as DataFrame.
- Automatic, display data alignment: display aligning objects to a set of labels, or ignoring labels, automatically aligning to data in Series, DataFrame calculations.
- Powerful, flexible grouping (group by) functions: split-apply-combine data sets, aggregate, transform data.
- Easy conversion of irregular, differently indexed data in Python and NumPy data structures into DataFrame objects.
- Slicing, fancy indexing, and subset decomposition of large datasets based on smart tags.
- Intuitive merging and joining of datasets, and the flexible reconstruction, pivoting of datasets.
- Axis support for structured tags: a scale supporting multiple tags.
- Mature IO tools: reading data from sources such as text files (CSV and other files supporting delimiters), Excel files, databases, etc., and saving/loading data using the ultra-fast HDF5 format.
- Time series: support date range generation, frequency conversion, moving window statistics, moving window linear regression, date displacement and other time series functions.

Pandas data structures are like containers for low-dimensional data. For example, DataFrame is a container for Series, and Series is a container for scalars. Using this approach, objects can be inserted or deleted in the container in the form of a dictionary.

In addition, the default operation of the generic API functions takes into account the orientation of the time series to the cross-sectional data set. When storing two- or three-dimensional data in multidimensional arrays, it is a burden to write functions that take into account the orientation of the dataset; in general, the different axes do not really make a difference in the program if the performance impact of continuity in C or Fortran is not taken into account. This is a "more appropriate" way to represent the direction of the data set. This allows us to use Pandas to write data conversion functions with less effort.

A particularly useful pandas feature for our research was that when dealing with tabular data such as DataFrame, index (rows) or columns (columns) are more intuitive than axis 0 and axis 1. Iterating over the columns of a DataFrame in this way makes the code easier to read and understand (see Figure 3 for an example)

```python
#drop columns that are not necssary for this project
cols_drop = ['CrimeCode','Total_Incidents','Post','Inside_Outside','Premise','RowID','X','Y']
bpd = bpd.drop(cols_drop, axis=1)
#Fill missing values in the District and Neighborhood with the mode and drop the remaining missing values
bpd['District'] = bpd['District'].fillna(bpd['District'].mode())
bpd['Neighborhood'] = bpd['Neighborhood'].fillna(bpd['Neighborhood'].mode())
bpd =bpd.dropna().reset_index(drop=True)
bpd.head()

    #check missing values
    def missing_value(df):
        """ Function to calculate the number and percent of missing values in a dataframe"""
        total = df.isnull().sum().sort_values(ascending=False)
        percent = ((df.isnull().sum()/df.isnull().count())*100).sort_values(ascending=False)
        missing_value = pd.concat([total, percent], axis=1, keys=['Total','Percent'])
        return missing_value
```

[Figure 3 image showing Pandas code and output table]

**Figure 3: Pandas Example**

## Method for Crime rate analysis

*Data Visualization:* Everything we touch seems to generate a lot of data. While buzzwords like "big data" may have disappeared, the data itself still exists. When it comes to understanding, processing, analyzing, and communicating this data, the old technology just won't cut it. It doesn't matter how big your screen is, or if it's curved, the CSV you import into Excel is too hard for anyone to understand.

Good visualizations help us get better output in less time. They also help us explain our work to colleagues, management and customers. These graphical representations can take many forms, from simple bar, column, line and pie charts; to dashboards containing multiple related graphs, to custom designed visualizations. Each type of chart has different advantages and disadvantages. Likewise, different charts are appropriate for different types of data.

The importance of data visualization cannot be overemphasized. There is no denying that humans are visual creatures. This is where data visualization comes in handy. It uses human perception and cognition to enhance understanding. Visualization uses the human ability to detect changes in size, shape, position, quantity and color. While we can see the exact numbers, it's often difficult to really understand what's going on looking only at numbers, and it's even harder to communicate a 'numbers-only' format with others within our organization. When viewing the same data in a visual way, the situation is completely different. We can immediately see some important details. Most importantly, we can easily communicate the data so we can investigate why this is happening and take the opportunity or resolve the issue with a team member. Now imagine the difference if you extrapolate to a more complex data set. In summary, data visualizations are important because they improve understanding by allowing humans to better process, analyze and communicate information.

*Introduction to Matplotlib:* Matplotlib is a visual manipulation interface for the Python programming language and its numerical mathematics extension, NumPy. It provides an application programming interface (API) to application embedded plots using common GUI toolkits such as Tkinter, wxPython, Qt or GTK+. In addition, matplotlib has a pylab interface based on image processing libraries (e.g., the open graphics library OpenGL) and is designed to be very similar to MATLAB-although perhaps not as user-friendly. sciPy uses matplotlib for graphical plotting.

The advantages of matplotlib.pylot are:
- Simple and easy to grasp for beginners.
- Easier to use for people who have had prior experience with Matlab or other graph plotting tools.
- It provides high-quality images and plots in various formats such as png, pdf, pgf, etc.
- Provides control to various elements of a figure such as DPI, figure colour, figure size.

*Introduction to Seaborn:* Seaborn is an advanced drawing library based on Matplotlib, with a higher level API encapsulated in the Matplotlib core library, which allows you to draw more beautiful graphics easily.

The advantages of seaborn are:
- Seaborn makes our charts and plots look engaging and enables some of the common data visualization needs (like mapping color to a variable or using faceting). Basically, it makes the data visualization and exploration easy to conquer. And trust me, that is no easy task in data science.
- Seaborn comes with a large number of high-level interfaces and customized themes that matplotlib lacks as it's not easy to figure out the settings that make plots attractive
- Matplotlib functions don't work well with dataframes, whereas seaborn does (see Figure 4 for an example).

```
In [1]: #import all needed libraries and perform any required environment setup
        import pandas as pd
        import calendar
        import squarify
        from pandas.tseries.holiday import USFederalHolidayCalendar as calender

        #expanding the number of visible colummns and rows
        pd.set_option('display.max_columns', 50)
        pd.set_option('display.max_rows', 500)

        import matplotlib.pyplot as plt
        import numpy as np
        import seaborn as sns
        import warnings
        warnings.filterwarnings('ignore')

        %matplotlib inline
        #configure the plotting style and size
        sns.set(style='darkgrid',palette='Dark2',rc={'figure.figsize':(9,6),'figure.dpi':100})
```

**Figure 4: Seaborn example**

*Introduction to TreeMap Chart:* Treemap Chart is intended for the visualization of hierarchical data in the form of nested rectangles. Each level of such a tree structure is depicted as a colored rectangle, often called a branch, which contains other rectangles (leaves). The space inside each of the rectangles that compose a Treemap is highlighted based on the quantitative value in the corresponding data point. See Figure 5 for an example.

The main advantages of Treemap Charts allow users to do the following:

- Identify the relationship between two elements in a hierarchical data structure.
- Optimize the use of space.

- Accurately display multiple elements together.
- Show ratios of each part to the whole.
- Visualize attributes by size and color coding.

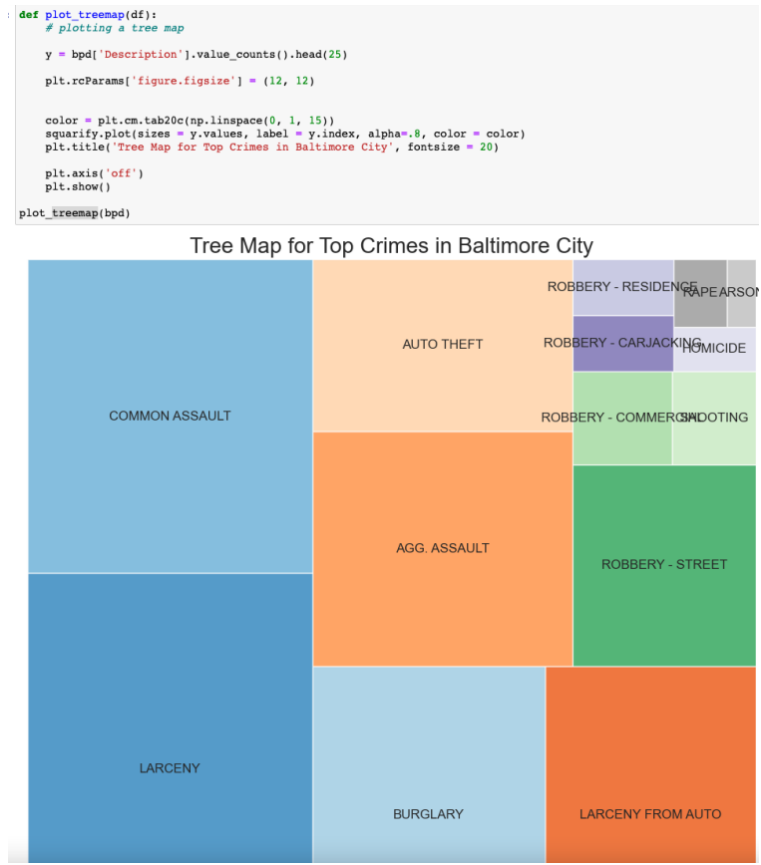

**Figure 5: Treemap Example**

Our team also leveraged several other methods, described briefly below.

*countplot*

- Show the counts of observations in each categorical bin using bars.
- A count plot can be thought of as a histogram across a categorical, instead of quantitative, variable. The basic API and options are identical to those for barplot(), so you can compare counts across nested variables.

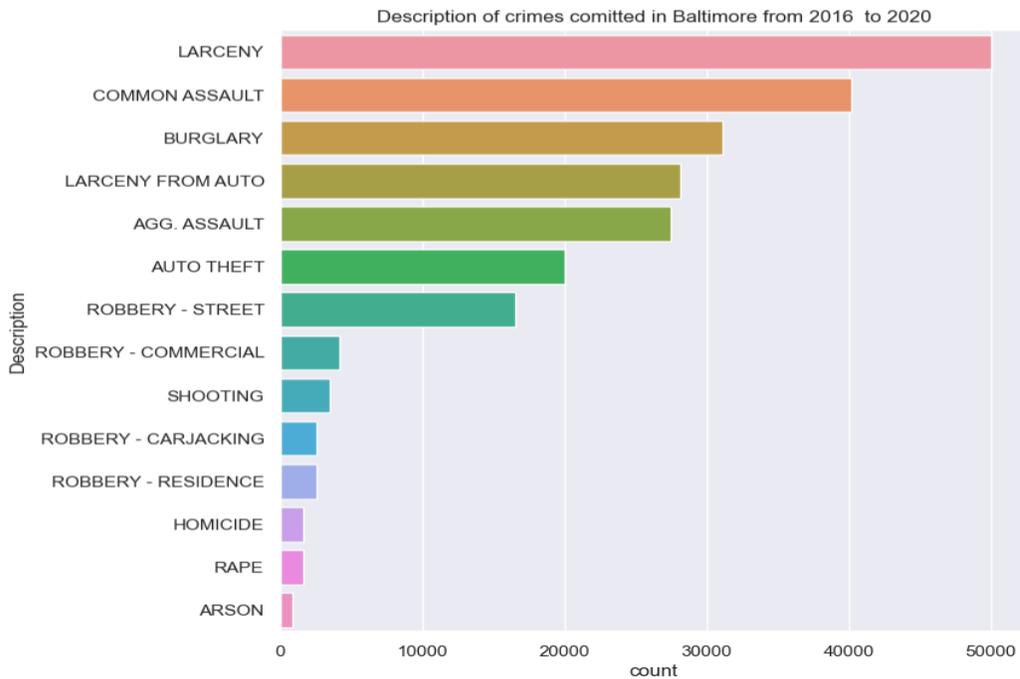

*Heatmaps*
- Giving direct overview of key web performance.
- Providing visual paths to understanding numeric values.
- Making it easier to learn from users to create more user friendly web design.
- Understand your visitors better, and provide them a better experience.

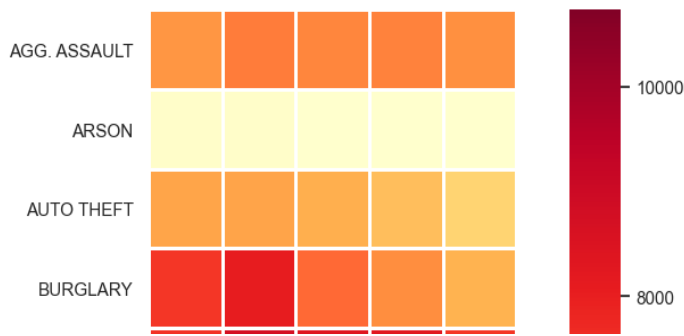

*PivotTables*
- Managing your data is easier with user-friendly features.
- Get valuable insights without hassle.
- Analyze data easier with Pivot Tables..
- Summarize data quickly and come up with decisions more efficiently with easily accessible insights.

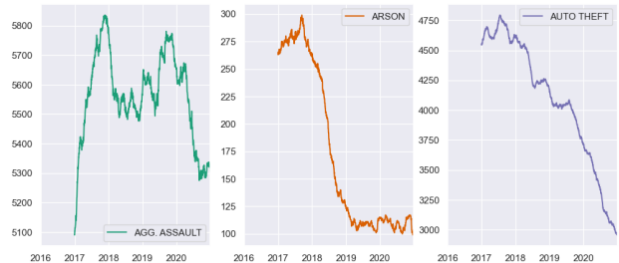

*pointplot:*
- This method is used to show point estimates and confidence intervals using scatter plot glyphs. A point plot represents an estimate of central tendency for a numeric variable by the position of scatter plot points and provides some indication of the uncertainty around that estimate using error bars.
- This function always treats one of the variables as categorical and draws data at ordinal positions (0, 1, … n) on the relevant axis, even when the data has a numeric or date type.

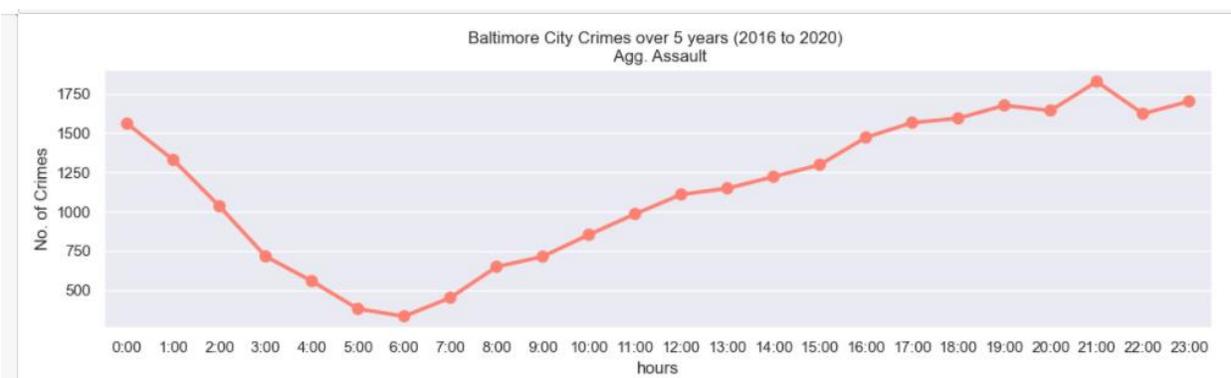

Pointplot also has the capability to combine charts.

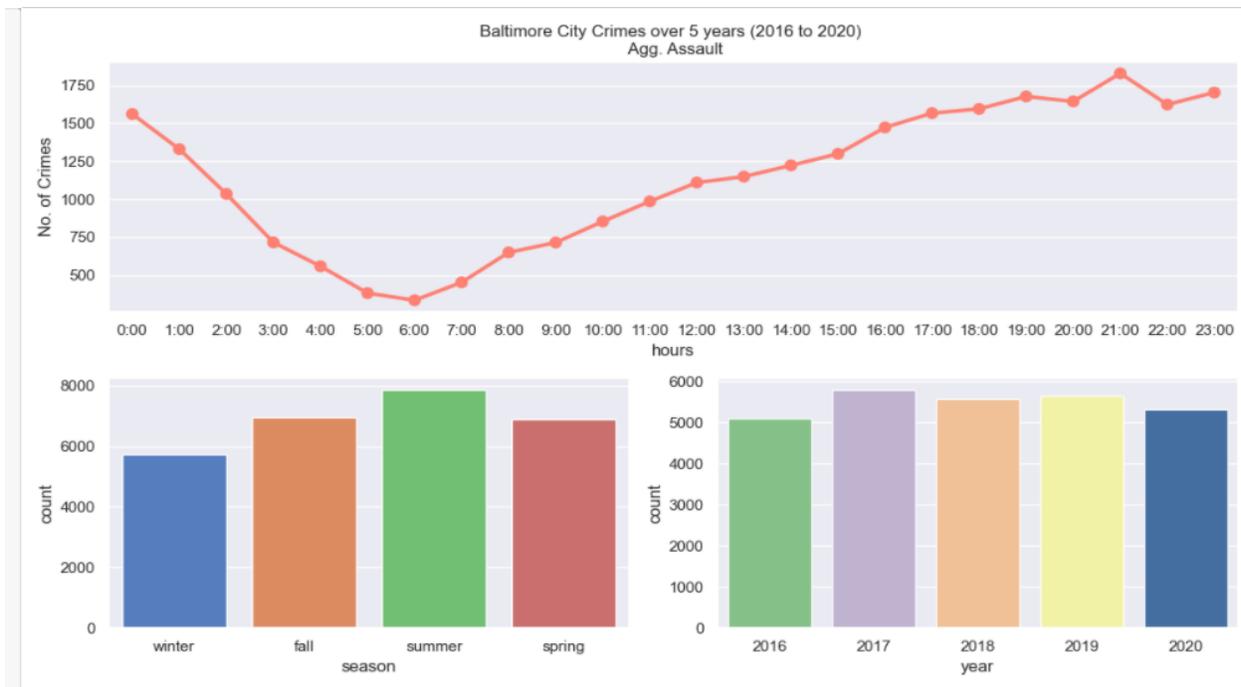

## Salary Data

This dataset includes Baltimore City employee salaries and gross pay from fiscal year 2011 through last fiscal year and includes employees who were employed on June 30 of the last fiscal year. For fiscal years 2020 and prior, data was extracted from the ADP payroll system. For fiscal year 2021, the data are combined from the ADP system and the Workday enterprise resource planning system which now includes payroll. See Figure 6 for an initial summary of the dataset.

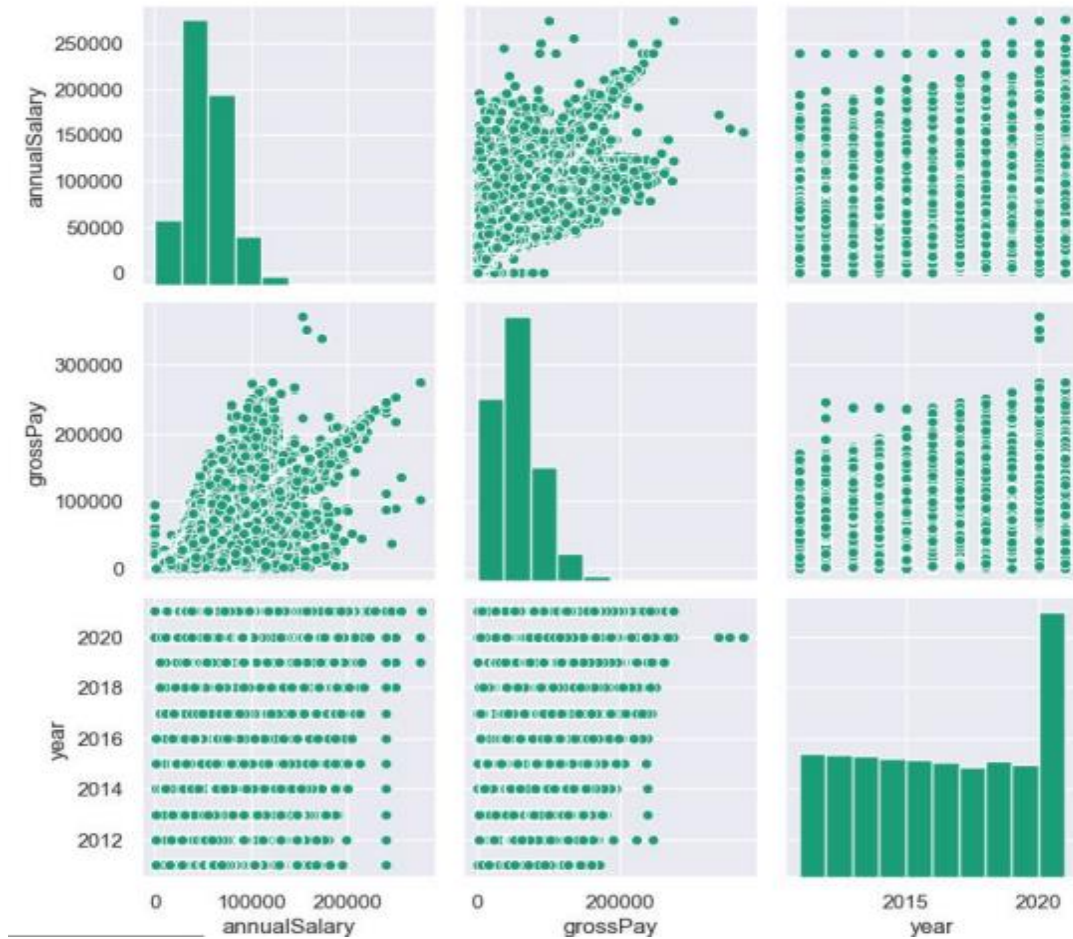

**Figure 6: Salary Data Summary**

There are around 1897 job classes and 71 agency IDs. Gross pay values are left skewed which indicates that the lowest gross pay values are more, skilled police officers alone get highly paid. There is an increase in gross pay and annual salary over the period of time which is expected. Gross pay and Annual salary goes hand in hand except in 2020.

## Linear Regression:

The purpose of regression is to predict, for example, tomorrow's weather temperature or the future the movement of stocks. Regression is predictive because it predicts future results from historical data. Linear regression is a concept originally used in statistics, but nowadays it is often used in machine learning. If there is a "linear relationship" between 2 or more variables, then we can use the historical data to figure out the "pattern" between the variables and build a valid model to predict the future outcome of the variables.

A primary advantage of linear regression is that the modeling is fast, does not require complex calculations, and runs very fast even with large amounts of data. The understanding and interpretation of each variable can be given based on the coefficients A main disadvantage is that it does not fit nonlinear data well. Hence it is necessary to determine whether the variables are linearly related to each other first.

Linear regression can model much more than linear relationships. 'Linearity' in linear regression refers to the linearity of the coefficients, while the functional relationship between the output and the features can be highly nonlinear through nonlinear transformations of the features and generalized linear models. On the other hand and more importantly, the ease of interpretation of the linear model makes it an irreplaceable model in the fields of physics, economics, and business.

## Data Facts:
*Crime data:* Analyzing 5 years (2016 to 2020) of Baltimore City crime data reveals that five crime types (Larceny, Common assault, Burglary, Larceny from auto, agg.assault) make up the majority of crime in the city, amounting to about 78% of the overall crime. Among the crime types reported, five (agg.assault, homicide, rape, robbery-carjacking and shooting) exhibit increasing trends with percentages increase that range from 8% to 400%. For most of these crime types, crime is low in the early morning hours from around 5am - 7am and rises steadily to its peak between 9pm and midnight. It also reveals that summer is the season when crime is likely to occur. Downtown is considered to be the most dangerous neighborhood because the crime count is highest. Rape on the other hand, usually occurs mostly around midnight and reduces gradually until it reaches the lowest in the early morning hours.

*Salary data:* We analyzed the salary data from years 2012 - 2020 and our analyzation clearly stated that average annual salary increased significantly from 2012 - 2020 which could be a standard rise in the salary adjustments(COLA) year by year, but the average gross pay (or overall compensation) did take a dip during the year 2020 and we highly suspect that there could be multiple factors which caused the dip, one of it could probably be Covid-19. However it did bounce back at the end of 2020 or early 2021 (see Figure 7).

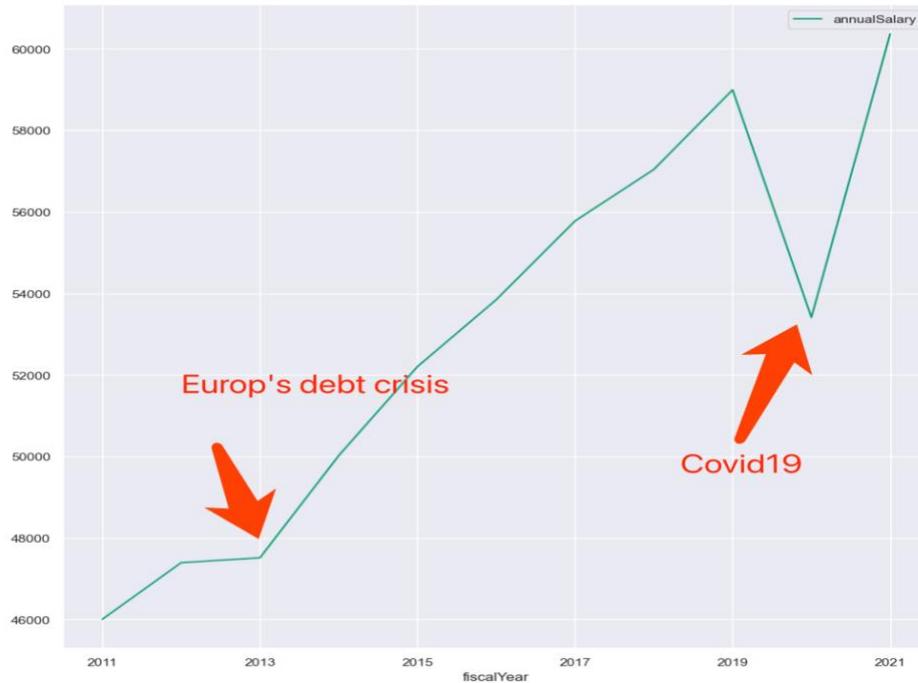

**Figure 7: Police Compensation Over Time.**

## Results: Crime Rates vs Salaries/Compensation:

Negative Correlation - when there is a negative correlation between two variables, the variables move relative to each other which means if one variable increases then the other decreases. When we analyzed the salary/compensation data from (2011-2021) we found that the overall wages were up, but we also saw that there was a relative decrease in compensation at some points and based on our research we sensed that this could have happened due to some historical events that occurred.

We used the Linear Regression Prediction model to compare the salary data with the crime data and we can clearly see that we obtained a negative correlated linear regression (see Figure 8).

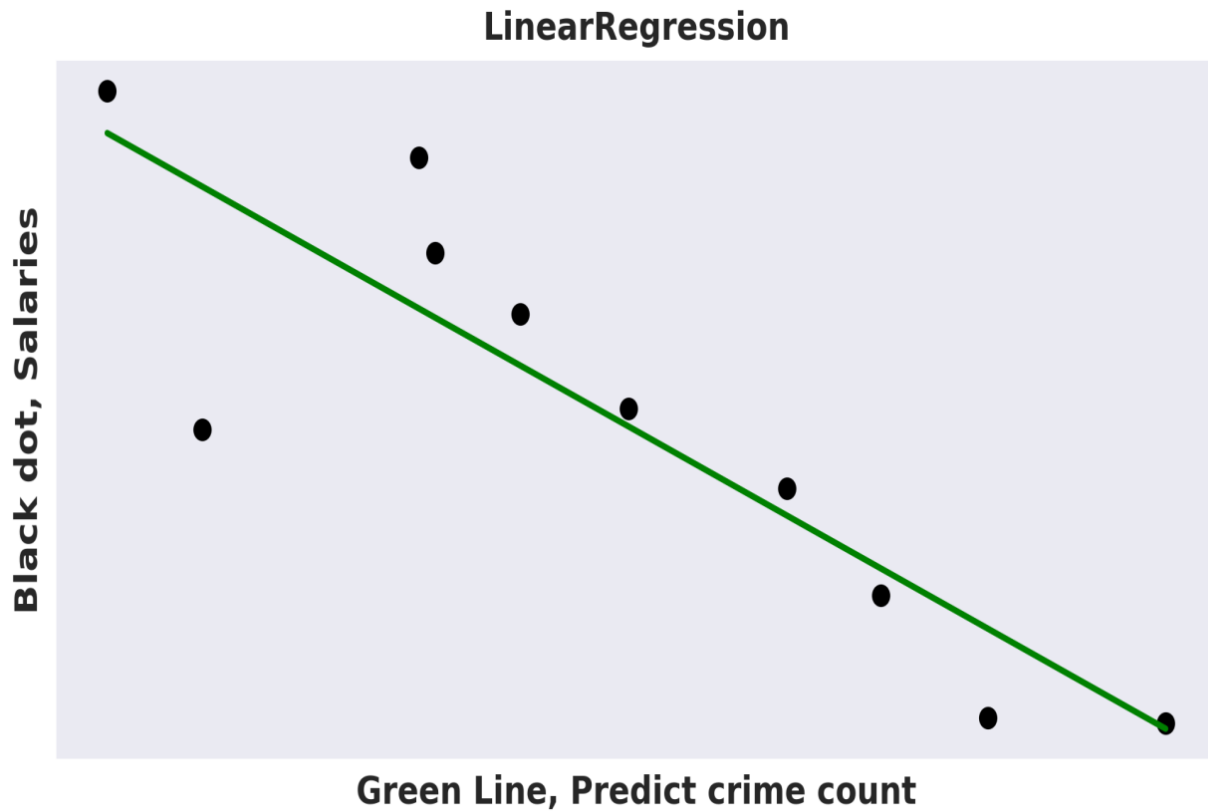

**Figure 8: Negative Correlation between Crime Rate and Police Compensation**

## Conclusion

Our initial results show a negative correlation between salary/compensation levels and crime rates. This is intuitive, in that one would expect to 'get what you pay for.' In other words, higher salaries are usually associated with greater competence and hence better outcomes. In this case, better police officers that can better enforce the laws and manage crime rates.

That said, the BPD is not a 'for profit' endeavor, and police officers are public servants that are compensated by acts of law, not via a more pure, capitalistic type of compensation method. Public Sector budgetary cycles tend to take much longer than the private sector, with a lot more bureaucratic 'friction.' Equally important, Police Officers tend to be very 'service-minded,' and not as motivated necessarily by compensation. Lastly, no one would advocate for paying police officers 'by the crime,' setting up a perverse, almost dystopian motivation scheme. For all these reasons, the link between salaries/compensation and crime rates is an interesting finding.

There is, however, a great deal more analysis that needs to make these findings actionable. For example, assessing the seniority mix that undergirds the salary data is important. Are the increased salaries a 'flight to quality,' if you will, where the BPD is made up of more senior police officers that are more effective, but earn a higher average salary? Or is it across the board higher compensation, including for junior police officers.

Moreover, this kind of analysis also leads to questions of interesting trade offs between a smaller, more senior police force with higher salaries (sort of a 'Special Ops' kind of concept), or is it better to use limited resources to simply have more police officers, including a significant junior (and compensation) officer corp. The literature is somewhat mixed on this question, with the same study suggesting that you need both: specialized units for high-risk repeat offenders, and extra police patrols for high crime spots[2]. There is also a massive repository of crime-policy related studies at the George Mason Center for Crime Based Policy that could be used to refine and apply the results of this research.

There is also some additional economic analysis that can be overlaid on these results to help bridge that gap to actual public policy. First and very simply, the salary analysis should be adjusted for inflation using the CPI index. A more in-depth analysis would leverage the work done by Gary Becker and William Landes[1], and provide a model for relating both police compensation and crime rate to the concept of 'Social Loss.' As shown in the formula below, according to Becker et. al., Social Loss (L) is a function of the damage done by crimes (D), the cost of combating offenses (C, of which police salary is a major driver), the cost of punishments (bf), and the 'activity level' (O, which is essentially the crime rate).

$$L = L(D, C, bf, O)$$

Using the formula above, our results could be extended to perhaps a multiple linear regression analysis that takes all of the four factors into account to or to minimize Social Loss. This may prove to be a useful tool for State Leadership to influence public policy, and to help decision makers in efforts to maximize the public good with limited resources.

8. Abrahamse, Allan F., Patricia A. Ebener, Peter W. Greenwood, Nora Fitzgerald, and Thomas E. Kosin (1991). "An Experimental Evaluation of the Phoenix Repeat Offender Program." Justice Quarterly 8:141–168.
9. Perfilyeva, A., Miskin, V. R., Aven, R., Drohan, C., & Ashqar, H. I. (2024). Estimating Variability in Hospital Charges: The Case of Cesarean Section. arXiv preprint arXiv:2411.08174.
10. https://governor.maryland.gov/2021/10/18/transcript-october-15-press-conference/
11. https://www.baltimorepolice.org/crime-stats/open-data
12. McCollister KE, French MT, Fang H. The cost of crime to society: new crime-specific estimates for policy and program evaluation. 2010 Apr 1; doi: 10.1016/j.drugalcdep.2009.12.002. Epub 2010 Jan 13. PMID: 20071107; PMCID: PMC2835847.
13. https://www.brennancenter.org/our-work/research-reports/predictive-policing-explained